\documentclass[twocolumn,english,aps,pra,showpacs]{revtex4}
\usepackage[T1]{fontenc}
\usepackage[latin1]{inputenc}
\usepackage{amsmath}
\usepackage{graphicx}
\usepackage{amssymb}
\usepackage{esint}

\begin{document}

\title{Spin Mixing in Spinor Fermi Gases}
\author{Ying Dong$^{1,2}$}
%\email{yingdong@mail.ustc.edu.cn}
\author{Han Pu$^{2,3}$}
%\email{hpu@rice.edu}
\affiliation{$^1$Department of Physics, Hangzhou Normal University, Hangzhou, Zhejiang 310036, China\\
$^2$Department of
Physics and Astronomy, and Rice Quantum Institute, Rice University, Houston, Texas 77251-1892, USA\\
$^3$Center for Cold Atom Physics, Chinese Academy of Sciences, Wuhan 430071, China}
\date{\today}

\begin{abstract}
We study a spinor fermionic system under the effect of spin-exchange interaction. We focus on the interplay between the spin-exchange interaction and the effective quadratic Zeeman shift. We examine the static and the dynamic properties of both two- and many-body system. We find that the spin-exchange interaction induces coherent Rabi oscillation in the two-body system, but the oscillation is quickly damped when the system is extended to the many-body case.
\end{abstract}


\maketitle
\section{Introduction}
Spinor quantum gas has received tremendous attention ever since the first creation of a spin-1 Bose-Einstein condensate in an optical trap \cite{dan}. Most of these studies deal with spinor Bose gases. The most salient feature of spinor condensates is the presence of spin-exchange interaction which drives coheret spin-mixing dynamics. A natural question can be asked is: can similar behavior be observed in a quantum degenerate fermionic system? The current work is an attempt to address this question.

Quantum degenerate Fermi gases have been realized in many laboratories \cite{Jin,Li,He,Yb1}. Different with the normal electronic system, many fermionic atoms have spins higher than 1/2 in their lowest hyperfine manifold. These large-spin ultracold Fermi gases provide us a unique opportunity to investigate exotic many-body physics \cite{color,1df1,1df2}, and have stimulated a great deal of theoretical interest \cite{Ho1,Ho2,Wu1,Wu2}. Considerable experimental progress has also been made recently in the system of $^{87}$Sr ($f=9/2$, where $f$ is total hyperfine spin) \cite{Sr} and $^{173}$Yb ($f=5/2$) \cite{Yb2}. Both $^{87}$Sr and $^{173}$Yb have an alkaline-earth-like atomic structure with all electron shells filled, thus their hyperfine spins completely come from nuclear spins. This will lead to a spin-independent atom-atom interaction since the nuclear spins are deep within the atom. In these alkaline-earth-like fermionic system, the spin-independent interaction can give rise to the so-called $SU(N)$ symmetry, with $N=2f+1$ \cite{sun1,sun2,sun3}. However, for a more general cold Fermi system including non-alkaline-earth atoms with large spins, the $SU(N)$ symmetry may not be reserved. As a result, one can find a more rich phases diagram in its ground state \cite{Wu3}. In a seminal work published very recently \cite{multi}, a spinor Fermi gas of $^{40}$K was realized. By taking advantage of the spin conservation, the effective spin of the system can be tuned from 1/2 to 9/2.

\begin{figure}
\includegraphics[width=2.5in]{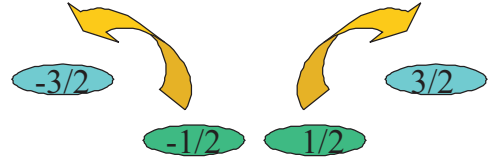}
\caption{ (Color Online) Schematic of a spin-mixing process.}
\label{mix}
\end{figure}

In this article, we focus on the simplest large-hyperfine-spin systems with $f=3/2$ with four internal components. This can be either true hyperfine spin (e.g., $^{132}$Cs, $^9$Be and $^{201}$Hg) or effetive spin such as realized in Ref.~\cite{multi}.
If we consider atoms with nonzero electron spins due to partially filled electron shells, then the interaction among atoms will be spin-dependent. One of the key features in this kind of systems is that there will be spin-exchange interactions which constantly mix different spin components. For example, two atoms with respective hyperfine spins $-1/2$ and $+1/2$ interact and become two atoms with hyperfine spins $-3/2$ and $+3/2$, as schemcatically depicted in Fig.~\ref{mix}. A similar case of spin mixing has been well studied for spinor condensates \cite{Pu1,Pu2,you,chang}.

The paper is organized as follows: We present the model Hamiltonian in Sec. II. The ground state properties and the spin-mixing dynamics are discussed in Secs. III and IV, respectively. Finally, we conclude in Sec. V.

\section{Model Hamiltonian}
%We first study the ground state of the system with the presence of an external magnetic field acting asymmetrically on the components of spin-$\pm1/2$ and spin-$\pm3/2$ respectively, Or in other words, the Zeeman shifts on components of spin-$\pm1/2$ and spin-$\pm3/2$ induced by the external magnetic field are unequal. We find that the ground state would be very different from our intuition under the large energy level difference limitation. Next, we simulate the internal dynamics of the spin-mixing process for the simplest two-body system, the Cooper pair, and the homogeneous many-body quantum gas. Our results qualitatively coincide with an experiment reported recently\cite{multi}.

%By solving the timedependent B-dG equations\cite{tdbdg}, We also simulate the internal dynamics of the spin-mixing process arising from the spin exchange interaction between atoms at a meanfield level in both BCS side and BEC side.

To begin we consider a homogeneous dilute gas of fermionic atoms with hyperfine spin $f=3/2$ in a box with volume $V$. The second quantized Hamiltonian of the system is given by
\begin{eqnarray}\label{ham1}
H&=&\int d{\bf r}[\sum_{\lambda}\psi_{\lambda}^{\dag}({\bf r})(-\frac{\hbar^2}{2m}\nabla^2+p_\lambda)\psi_{\lambda}({\bf r})\\ \nonumber
&&+\sum_{\lambda_1,\lambda_2,\lambda_3,\lambda_4}U_{\lambda_1\lambda_2\lambda_3\lambda_4}\psi_{\lambda_1}^{\dag}({\bf r})\psi_{\lambda_2}^{\dag}({\bf r})\psi_{\lambda_3}({\bf r})\psi_{\lambda_4}({\bf r})],
\end{eqnarray}
where $\psi_\lambda$ $(\lambda=-3/2,-1/2,1/2,3/2)$ is the atomic field annihilation operator associated with atoms in the hyperfine spin state $|f=3/2,m_f=\lambda\rangle$. The summation indices in (\ref{ham1}) run through the values $-3/2,-1/2,1/2,3/2$. $p_\lambda$ is the bare atomic energy for spin state $\lambda$. We will consider an effective quadratic Zeeman shift such that $p_{-1/2}=p_{1/2}$ and $p_{-3/2}=p_{3/2}$ as the linear Zeeman shift can be gauged away. If we consider the $s$-wave scattering only, then the interaction between atoms can be characterized by the coefficients $U_{\lambda_1\lambda_2\lambda_3\lambda_4}$ which are obtained from the two-body interaction model
$\hat{U}=g_0\hat{P}_0+g_2\hat{P}_2$.
Here, $P_F$ is the projection operator which projects the pair into a total hyperfine spin $F$ state, $g_F$ is the interaction strength in the total spin $F$ channel, which, in the calculation, will be replaced
in favor of the $s$-wave scattering length $a_F$ via the regularization
procedure:
\begin{equation}
\frac{1}{g_F} \to \frac{m}{4\pi\hbar^2a_F}-\frac{1}{V}\sum_{\bf k}\frac{m}{\hbar^2{\bf k}^2}\,.
\end{equation}
For Fermi gases, there is no $s$-wave interaction in the odd total spin ($F=1$, 3) channels, since these channels are forbidden by Pauli's exclusion principle.
%Generally, these interactions are described by two parameters in the total spin $F=0,2$ channels as $g_{0,2}=4\pi\hbar^2a_{0,2}/m$ with $a_{0,2}$ the corresponding $s$-wave scattering lengths and $m$ the atom mass.
%\begin{eqnarray}
%H&=&\int d {\bf r}[\sum_{\lambda}\psi_{\lambda}^{\dag}(-\frac{\hbar^2}{2m}\nabla^2-\mu)\psi_{\lambda}+
%\frac{g_0+g_2}{2}(\psi_{\alpha}^{\dag}\psi_{\beta}^{\dag}\psi_{\beta}\psi_{\alpha}+\psi_{\mu}^{\dag}\psi_{\nu}^{\dag}\psi_{\nu}\psi_{\mu})
%+\frac{g_0-g_2}{2}(\psi_{\alpha}^{\dag}\psi_{\beta}^{\dag}\psi_{\nu}\psi_{\mu}+\psi_{\mu}^{\dag}\psi_{\nu}^{\dag}\psi_{\beta}\psi_{\alpha}) %\nonumber \\
%& %&{}+g_2(\psi_{\alpha}^{\dag}\psi_{\mu}^{\dag}\psi_{\mu}\psi_{\alpha}+\psi_{\alpha}^{\dag}\psi_{\nu}^{\dag}\psi_{\nu}\psi_{\alpha}+\psi_{\beta}^{\dag}\psi_{\mu}^{\dag}\psi_{\mu}\psi_{\beta}+\psi_{\beta}^{\dag}\psi_{\nu}^{\dag}\psi_{\nu}\psi_{\beta})]
%\end{eqnarray}

We expand the field operators with plane wave function
$\psi_{\lambda}=\sum_{{\bf k}} \lambda_{{\bf k}}e^{i{\bf k}\cdot{\bf r}}/\sqrt{V}$
and complete the spatial integral to the Hamiltonian into momentum space:
\begin{widetext}
\begin{eqnarray}
H&=&\sum_{{\bf k},\lambda}\mathcal{E}_{\lambda {\bf k}} \lambda^{\dag}_{{\bf k}}\lambda_{{\bf k}}+\sum_{{\bf k},{\bf k}^{\prime},{\bf p}}[
c_2(\alpha^{\dag}_{{\bf k}^{\prime}+{\bf p}}\beta^{\dag}_{{\bf k}-{\bf p}}\nu_{{\bf k}}\mu_{{\bf k}^{\prime}}+\mu^{\dag}_{{\bf k}^{\prime}+{\bf p}}\nu^{\dag}_{{\bf k}-{\bf p}}\beta_{{\bf k}}\alpha_{{\bf k}^{\prime}})
+c_0(\alpha^{\dag}_{{\bf k}^{\prime}+{\bf k}}\beta^{\dag}_{{\bf k}-{\bf p}}\beta_{{\bf k}}\alpha_{{\bf k}^{\prime}}+\mu^{\dag}_{{\bf k}^{\prime}+{\bf p}}\nu^{\dag}_{{\bf k}-{\bf p}}\nu_{{\bf k}}\mu_{{\bf k}^{\prime}})\nonumber\\
& &{}+\frac{g_2}{V} \,(\alpha^{\dag}_{{\bf k}^{\prime}+{\bf p}}\mu^{\dag}_{{\bf k}-{\bf p}}\mu_{{\bf k}}\alpha_{{\bf k}^{\prime}}+\beta^{\dag}_{{\bf k}^{\prime}+{\bf p}}\nu^{\dag}_{{\bf k}-{\bf p}}\nu_{{\bf k}}\beta_{{\bf k}^{\prime}}+\alpha^{\dag}_{{\bf k}^{\prime}+{\bf p}}\nu^{\dag}_
{{\bf k}-{\bf p}}\nu_{{\bf k}}\alpha_{{\bf k}^{\prime}}+\mu^{\dag}_{{\bf k}^{\prime}+{\bf p}}\beta^{\dag}_{{\bf k}-{\bf p}}\beta_{{\bf k}}\mu_{{\bf k}^{\prime}})]\,, \label{H}
\end{eqnarray}
\end{widetext}
Here, $c_0=(g_0+g_2)/{2V}$, $c_2=(g_2-g_0)/{2V}$ and $\mathcal{E}_{\lambda {\bf k}}=\hbar^2k^2/2m+p_\lambda$. For notational simplicity, we have used $\alpha$, $\beta$, $\mu$ and $\nu$ to denote the annihilation operators for the component with spin quantum number $m_f=1/2,\;-1/2,\;3/2 $ and $-3/2$, respectively.
From the Hamiltonian above, we can see clearly that the terms with coefficients $c_2$ describe spin-mixing processes such as that depicted in Fig.~\ref{mix}. Thus spin mixing requires $c_2 \neq 0$ or $g_0 \neq g_2$. In the case $g_0 = g_2$, spin mixing does not exist and the population in each spin component is individually conserved. Under such a condition, the interaction obeys $SU(4)$ symmetry \cite{Wu1,Wu2}.

As our focus is on the effect of spin-mixing interaction, we will consider the case with $g_0 \neq g_2$. Great simplification can be further achieved by assuming $g_2=0$ and $g_0 \neq 0$, in which case the second line of Eq.~(\ref{H}) vanishes. This is the case we will consider in this work. We note that the essential physics does not change qualitatively if $g_2 \neq 0$.

%\section{R=0}
\section{Ground State Properties}
Throughout this work, we take temperature to be zero. We will first consider the mean-field ground state property of the system. In the case of $g_2=0$, it is obvious from Hamiltonian (\ref{H}) that superfluid pairing can occur between spin components $(\frac{3}{2},-\frac{3}{2})$ and $(\frac{1}{2},-\frac{1}{2})$. If we denote
%\begin{equation}
%H=\sum_{\vec{k},\lambda}\mathcal{E}_k \lambda^{\dag}_{\vec{k}}\lambda_{\vec{k}}+\sum_{\vec{k}\vec{k}^{\prime}}[
%E(\alpha^{\dag}_{\vec{k}^{\prime}}\beta^{\dag}_{-\vec{k}^{\prime}}\nu_{-\vec{k}}\mu_{\vec{k}}+ %\mu^{\dag}_{\vec{k}^{\prime}}\nu^{\dag}_{-\vec{k}^{\prime}}\beta_{-\vec{k}}\alpha_{\vec{k}})+D(\alpha^{\dag}_{\vec{k}^{\prime}}\beta^{\dag}_{-\vec{k}^{\prime}}\beta_{-\vec{k}}\alpha_{\vec{k}}+ %\mu^{\dag}_{\vec{k}^{\prime}}\nu^{\dag}_{-\vec{k}^{\prime}}\nu_{-\vec{k}}\mu_{\vec{k}})]
%\end{equation}
$C_1=\sum_{\bf k}\langle \beta_{-\bf k}\alpha_{\bf k}\rangle$ and $C_3=\sum_{\bf k}\langle \nu_{-\bf k}\mu_{\bf k} \rangle$,
the Hamiltonian under the mean-field approximation can be written as
\begin{eqnarray}
H  &=&  \sum_{{\bf k},\lambda} (\mathcal{E}_{\lambda{\bf k}}+h_{\lambda})\lambda_{\bf k}^{\dagger}\lambda_{\bf k} \nonumber \\
&&+\sum_{\bf k} [
\Delta^{\ast}\beta_{-\bf k}\alpha_{\bf k}-\Delta^{\ast}\nu_{-\bf k}\mu_{\bf k}+h.c.]\,,
\end{eqnarray}
where $h_\lambda=g_0\sum_{\bf k} \langle \lambda_{\bf k}^\dagger\lambda_{\bf k}\rangle /2V$ and $\Delta=\frac{g_0}{2V}(C_1-C_3)$.
To diagonalize this Hamiltonian, we can perform the Bogoliubov transformation
\begin{eqnarray}
\alpha_{\bf k}=u_{\bf k}a_{\bf k}+v_{\bf k}b^{\dag}_{\bf -k},&\quad &\beta_{\bf -k}=u_{\bf k}b_{\bf -k}-v_{\bf k}a_{\bf k}^{\dag}\nonumber\\
\mu_{\bf k}=s_{\bf k}u_{\bf k}+t_{\bf k}v^{\dag}_{\bf -k},&\quad& \nu_{\bf -k}=s_{\bf k}v_{\bf -k}-t_{\bf k}u_{\bf k}^{\dag}
\end{eqnarray}
Here, the new coefficients  should meet the following conditions to ensure the anticommutativity of new operators.
\begin{equation}
|s_{\bf k}|^2+|t_{\bf k}|^2=1,\quad {\rm and} \quad |u_{\bf k}|^2+|v_{\bf k}|^2=1 \,.
\end{equation}
Then we can write down the Bogoliubov-de Gennes (BdG) equations as
\begin{eqnarray}\label{bdg}
\left(\begin{array}{cc}
\mathcal{E}_{\alpha {\bf k}}+h_{\alpha}-\tilde{\mu} & \Delta\\
\Delta^{\ast}&-\mathcal{E}_{\beta{\bf k}}-h_\beta+\tilde{\mu}\\
\end{array}
\right)\left(\begin{array}{c}
u_{\bf k}\\
v_{\bf k}
\end{array}
\right)=E_1\left(\begin{array}{c}
u_{\bf k}\\
v_{\bf k}
\end{array}
\right)\nonumber
\end{eqnarray}
\begin{eqnarray}\label{bdg}
\left(\begin{array}{cc}
\mathcal{E}_{\mu{\bf k}}+h_\mu-\tilde{\mu} &-\Delta\\
-\Delta^{\ast}&-\mathcal{E}_{\nu{\bf k}}-h_\nu+\tilde{\mu} \\
\end{array}
\right)\left(\begin{array}{c}
s_{\bf k}\\
t_{\bf k}
\end{array}
\right)=E_3\left(\begin{array}{c}
s_{\bf k}\\
t_{\bf k}
\end{array}
\right)
\end{eqnarray}
where we have introduced the chemical potential $\tilde{\mu}$ into the equations.
We assume that there are $2N$ particles in total, and the population in opposite spin states are equal, i.e., $N_\alpha=N_\beta$ and $N_\mu=N_\nu$. Then we have $ N=\sum_{\bf k}(|v_{\bf k}|^2+|t_{\bf k}|^2)$ and the order parameter is given by
\begin{equation}
\Delta=-\frac{g_0}{2V}\sum_{\bf k}(u_{\bf k}v_{\bf k}^{\ast}-s_{\bf k}t_{\bf k}^{\ast}) \,.
\end{equation}

\begin{figure}
\includegraphics[width=3.5in]{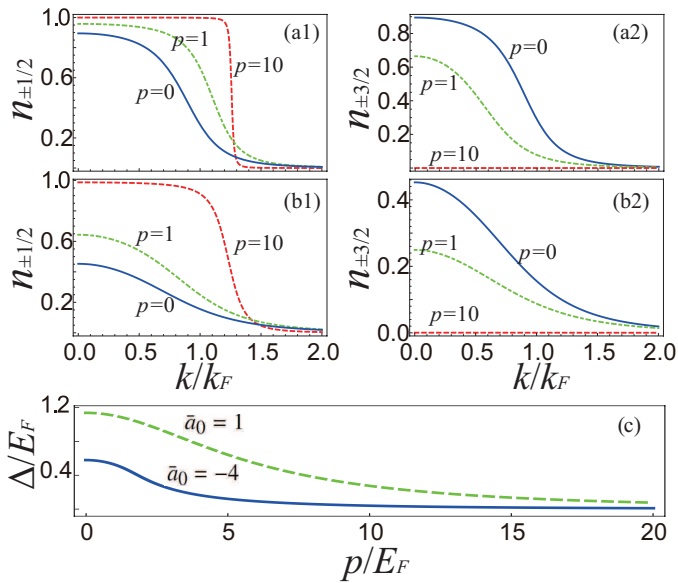}
\caption{ (Color Online) (a), (b) The momentum distribution in spin components $\pm 1/2$ and $\pm 3/2$ for quadratic Zeeman shift $p=0$ (solid blue), $p=1$ (dot-dashed orange) and $p=10$ (dashed red), in units of the Fermi energy $E_F$. In the calculation we set $\bar{a}_0=-4$ for (a1) and (a2), and $\bar{a}_0=1$ for (b1) and (b2), where the dimensionless interaction strength is defined as $\bar{a}_0=16 a_0 k_F/\pi$ with $k_F$ being the Fermi wavenumber. (c) Order parameter $\Delta$ as a function of $p$.}
\label{nk}
\end{figure}

Without loss of generality, we take the effective quadratic Zeeman shift as $p_\alpha=p_\beta=p_{\pm 1/2}=0$ and $p_\mu = p_\nu=p_{\pm 3/2}=p \ge 0$. For a given $p$ and the $s$-wave scattering length $a_0$, we solve the BdG equations self-consistently, and typical results are presented in Fig.~\ref{nk}.

The upper panels of Fig.~\ref{nk} display the momentum distribution in different spin components at various values of $p$. As $p$ increases, the population in the spin components $m_f=\pm 3/2$ decreases, and that in $m_f=\pm 1/2$ increases. This can be easily understood from the energetic point of view. However, somewhat surprisingly, for large $p$, the momentum distribution in $m_f=\pm 1/2$ components approaches a step function, exemplifying a normal Fermi sea. Indeed, Fig.~\ref{nk}(c) shows that the magnitude of the order parameter decreases as $p$ increases. Intuitively, one might have thought that for very large $p$, the population in the $m_f=\pm 3/2$ components becomes negligible and the system reduces into a two-component spin-1/2 Fermi gas, which should exhibits a superfluid ground state at zero temperature. Results presented in Fig.~\ref{nk} apparently contradicts this intuitive picture.

To gain some insights into this problem, let us consider the so-called Cooper problem where two extra particles with attractive interaction lie on the surface of a filled Fermi sea noninteracting atoms. The eigenvalue equations describing the two interacting particles in ${\bf k}$-space are given by
\begin{eqnarray}\label{eigenvalue}
E\psi_{\alpha\beta}({\bf k})\!\!&=&\!\! E_{\alpha\beta}\psi_{\alpha\beta}({\bf k})+\frac{g_0}{2V}\sum_{k^{\prime}}\left(\psi_{\alpha\beta}({\bf k^{\prime}})-\psi_{\mu\nu}({\bf k^{\prime}})\right) \nonumber \\
E\psi_{\mu\nu}({\bf k})\!\! &=& \!\! E_{\mu\nu}\psi_{\mu\nu}({\bf k})+\frac{g_0}{2V}\sum_{k^{\prime}}\left(\psi_{\mu\nu}({\bf k^{\prime}})-\psi_{\alpha\beta}({\bf k^{\prime}})\right)
\end{eqnarray}
where $E_{\alpha\beta}=\frac{\hbar^2k^2}{m}-2E_{F}$ and $E_{\mu\nu}=\frac{\hbar^2k^2}{m}-2E_{F}+2p$, and $\psi_{\alpha \beta}$ and $\psi_{\mu \nu}$ are pairing amplitudes between $m_f= \pm 1/2$, $m_f=\pm 3/2$, respectively. The presence of the Fermi sea imposes the restriction that ${\bf k}$ and ${\bf k}'$ lie outside of the Fermi sea, i.e., $k, k' \ge k_F$. A negative value of the eigenenergy $E$ means that the two particles form a bound pairing with binding energy $|E|$. The larger $|E|$ is, the more strongly the pair binds.

\begin{figure}
\includegraphics[width=2.7in]{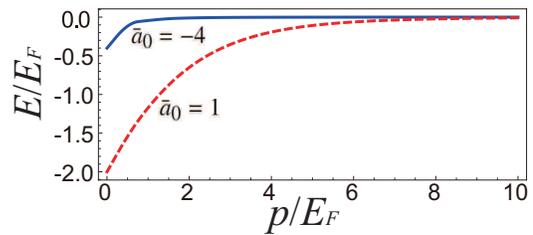}
\caption{ (Color Online) The bound state energy of the Cooper pair as a function of $p$ for dimensionless interaction strength (defined in Fig.~\ref{nk}) $\bar{a}_0=-4$ (solid blue) and $\bar{a}_0=1$ (dashed red), respectively.  }
\label{cooper}
\end{figure}
After some mathematical manipulation, the two eigenvalue equations (\ref{eigenvalue}) leads to the following single equation for $E$:
\begin{equation}
1=\frac{g_0}{2V}\,\sum_{k>k_F}\left(\frac{1}{E-E_{\alpha \beta}}+\frac{1}{E-E_{\mu \nu}} \right)\,,
\end{equation}
which can be solved numerically. The solution of $E$ as a function of $p$ is displayed in Fig.~\ref{cooper}, from which one can see that, as $p$ increases, $|E|$ tends to zero. In other words, the pairs becomes less and less bound. This result is consistent with the many-body result illustrated in Fig.~\ref{nk}.
%Converting the summation to an integral by introducing a suitable density of states and integrate it out, we will get the eigenenergy $E$ as a function of $p$. From Fig.2, we can see that as $p$ becomes larger and larger, the eigenenergy $E$ tends to zero rapidly in both BCS side and BEC side. This means that when the Zeeman shift on spin state $|m_f=\pm\frac{3}{2}\rangle$ induced by external magnetic field is much larger than $E_\mathcal{F}$, the two atoms outside the Fermi sea will tend to stay freely rather than form a Cooper pair.

We therefore reach the conclusion that the energy mismatch induced by the quadratic Zeeman shift $p$, together with the spin-exchange interaction, tends to break the pair apart. This phenomenon is reminiscent of the effect by a magnetic impurity on a spin-1/2 superconductor \cite{zhu06,lei}. In this latter system, the magnetic impurity induces an energy difference between the two pairing particles and has the tendency of destroying the pairing.
%
%In fact, it is not difficult to understand this effect in our model with spin-exchange interaction. If the two atoms outside the Fermi sea form a Cooper pair, then it would be possible for these two atoms to stay on a large momentum state, that is $|\psi_{\pm1/2}(k)|\ne 0$ for large $k$. Two atoms with high energy in in spin states $|\pm\frac{1}{2}\rangle$ can transit to spin states $|\pm\frac{3}{2}\rangle$ under the spin-exchange interaction. This transition will increase the energy of the whole system since the spin states $|\pm\frac{3}{2}\rangle$ have a much higher energy level. In the regime of large $p$, even a very small tail of large momentum can increase the average energy of the whole system dramatically. So, in ground state, to minimize the whole energy, the best choice for the two atoms is that they should stay as close as possible to the surface of the Fermi sea.

\section{Spin-Mixing Dynamics}
So far we have focused on the ground state of the system.
Now, let us turn to the spin-mixing dynamics. Before dealing with the many-body situation, it may be
helpful to investigate the Cooper problem first. We take the initial state to be the ground state of the Cooper system under an effective quadratic Zeeman field with $p=10 E_F$. At $t=0$, we suddenly turn this field off so that $p=0$ and the system starts to evolve. The dynamics of system is governed by Eq.~(\ref{eigenvalue}) after replacing $E$ on the left hand side with $i\hbar\partial / \partial t$. The equations can be simplified if we redefine two quantities as follows:
\[ \psi_\pm ({\bf k},t) = \frac{1}{\sqrt{2}} \,[ \psi_{\alpha \beta}  ({\bf k},t) \pm \psi_{\mu \nu}  ({\bf k},t) ] \,,\] and the dynamical equations for $p=0$ can be rewritten as
\begin{eqnarray}
i\hbar \frac{\partial}{\partial t} \psi_+  ({\bf k},t) &=& \frac{\hbar^2 k^2}{m}  \psi_+  ({\bf k},t)  \,,\\
i\hbar \frac{\partial}{\partial t} \psi_-  ({\bf k},t) &=& \frac{\hbar^2 k^2}{m}  \psi_-  ({\bf k},t)  + \frac{g_0}{V} \sum_{{\bf k}'} \psi_-  ({\bf k}',t)\,.
\end{eqnarray}
Hence the two equations for $\psi_{\pm}$ are decoupled and the interaction term only appears in the equation for $\psi_-$.

\begin{figure}
\includegraphics[width=3.in]{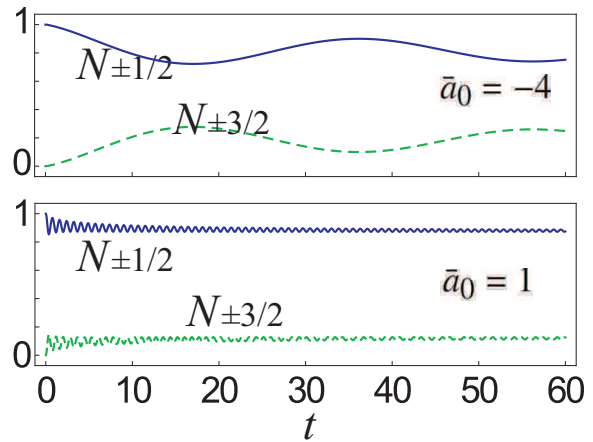}
\caption{(Color Online) Populations in spin states $m_f=\pm 1/2$ and $m_f=\pm 3/2$  as functions of time (in units of $\hbar/E_F$) at two different interaction strengths.}
\label{Cd}
\end{figure}

The above equations can be easily solved and results are presented in Fig.~\ref{Cd}. Initially, due to the presence of the large quadratic Zeeman shift, almost all the populations are in spin states $m_f=\pm 1/2$. Spin mixing dynamics is initiated by quenching this Zeeman shift at $t=0$. We compare the cases for two different values of the interaction strength: a weak interaction with $\bar{a}_0=-4$ and a strong interaction with $\bar{a}_0=1$. In both scenarios, damping is observed in spin mixing, and stronger interaction gives rise to a much faster damping. This can be intuitively understood as follows: The initial state can be regarded as a superposition of different eigenstates of the quenched Hamiltonian. The stronger the interaction, the larger the number of the eigenstates contained in the initial state. For $t>0$, different eigenstates oscillate at different frequencies which results in the damping of the population dynamics. The more eigenstates are involved, the faster the damping. Therefore, such damping is a result of the intrinsic multi-mode nature of the Fermi gas. For a system of spinor condensate near zero temperature, as all the atoms occupy the same lowest-energy orbitals, nearly undamped spin-mixing oscillations can be observed \cite{Pu2,you, chang}.

We now turn our attention to the dynamics in the many-body setting. As in the Cooper problem, we prepare the system in the ground state with $p=10 E_F$ and at $t=0$ quench the quadratic Zeeman field to zero. The ensuing mean-field dynamics can be simulated by solving the time-dependent BdG equations, obtained by replacing the eigenenergies at the right hand side of
Eqs.~(\ref{bdg}) with $i\hbar \partial /\partial t$ \cite{andrew}. At each time step, the order parameter will be updated as  \[ \Delta(t)=-\frac{g_0}{2V}\sum_{\bf k}(u_{\bf k}(t)v_{\bf k}^{\ast}(t)-s_{\bf k}(t)t_{\bf k}^{\ast}(t)) \,.\]

\begin{figure}
\includegraphics[width=3.in]{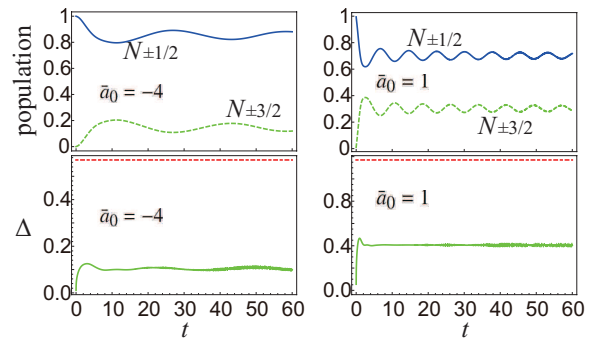}
\caption{ (Color Online) Time dependence of population (upper panels) and order parameter $\Delta$ (lower panels), in units of $E_F$, for two different interaction strengths. The horizontal lines in the lower panels represent the value of $\Delta$ in the ground state with $p=0$. Time is in units of $\hbar /E_F$.}
\label{mixmany}
\end{figure}

Representative results are shown in Fig.~\ref{mixmany}. The dynamics exhibits qualitatively similar properties as in the Cooper problem: damping is observed in both the dyanmics of the population and that of the order parameter, and the stronger the interaction, the faster the damping. This can be understood using a similar intuitive argument we presented for the Cooper problem. We note that such damping was observed in the experiment reported in Ref.~\cite{multi}. Finally we remark that the ground state corresponding to $p=0$ should have equal population in all four spin states and an order parameter value indicated by the horizontal line in the lower panels of Fig.~\ref{mixmany}. In the absence of any dissipation, as in our simulation, the system remains far away from the ground state.

%\begin{figure}
%\includegraphics[width=3.5in]{bcs-mix.eps}
%\caption{ (Color Online)Time dependence of particle number for initial states prepared in BCS states restricted in %components of spin $\pm1/2$. The other parameters are all same with Fig.3. }
%\end{figure}

%\begin{figure}
%\includegraphics[width=3.5in]{delta.eps}
%\caption{ (Color Online) Time dependence of gap value(normalized by $E_\mathcal{F}$) for $\bar{a}_0=-4$(BCS side) and $\bar{a}_0=1$(BEC side) respectively. The green line(dotdashed) and the blue line(dashed) stand for the dynamics with a initial normal-like state and BCS state restricted in components of spin $\pm1/2$ respectively. The red line (solid) stands for the gap energy of the real ground states for the corresponding parameters.}
%\end{figure}

\section{Conclusion}
To conclude, we have examined the spin mixing interaction of a degenerate Fermi gas with four internal spin components. It is quite remarkable that in the presence of an effective quadratic Zeeman field that shifts relatively the bare energies of spin-($\pm 1/2$) and spin-($\pm 3/2$) states, the system becomes almost normal. This is analogous to the effect of the Zeeman field that breaks the symmetry of the two spin components in a two-component (spin-1/2) Fermi gas. We have also investigated the spin-mixing dynamics initiated by quenching the quadratic Zeeman field and show that, unlike in the case of a spinor condensate, damping will necessarily occur in a many-body Fermi gas due to the intrinsic multi-mode nature of fermions.
%In the future, we hope to address this question as well as to generalize our model from the homogeneous system to a harmonic trapped system which is much closer to the experimental circumstance.

\begin{acknowledgments}
This work is supported by the ARO Grant
No. W911NF-07-1-0464 with the funds from the DARPA
OLE Program, the Welch foundation (C-1669, C-1681),
and the NSF. Y. Dong acknowledges financial support from NNSF of China (Grant No. 11274085).
\end{acknowledgments}

\end{document}